\documentclass[12pt]{article}

\usepackage{newtxtext,newtxmath}

\usepackage{graphicx}

\usepackage[letterpaper,margin=1in]{geometry}

\renewenvironment{abstract}
	{\quotation}
	{\endquotation}

\date{}

\makeatletter
\renewcommand{\fnum@figure}{\textbf{Figure \thefigure}}
\renewcommand{\fnum@table}{\textbf{Table \thetable}}
\makeatother

\usepackage{scicite}

\usepackage{url}

\def\scititle{
	Hybrid Implementation for Untrusted-node-based Quantum Key Distribution Network
}
\title{\bfseries \boldmath \scititle}

\author{
	Jingyang Liu$^{1,2,3}$,
	Xingyu Zhou$^{1,2,3}$,
        Huajian Ding$^{1,2,3}$,
        Jiaxin Xu$^{1,2,3}$, \\
        Chunhui Zhang$^{1,2,3}$,
        Jian Li$^{1,2,3}$,
	and Qin Wang$^{1,2,3*}$\\
	\small$^{1}$Institute of quantum information and technology, \\ \small Nanjing University of Posts and Telecommunications, Nanjing 210003, China.\and
	\small$^{2}$“Broadband Wireless Communication and Sensor Network Technology”Key Lab of Ministry of Education, \\ \small Ministry of Education, Nanjing 210003, China.\and
        \small$^{3}$“Telecommunication and Networks”National Engineering Research Center, NUPT, Nanjing 210003, China\and
	\small$^\ast$Corresponding author. Email: qinw@njupt.edu.cn\and
}

\begin{document} 

\maketitle

\begin{abstract} \bfseries \boldmath
Quantum key distribution (QKD) serves as a cornerstone of secure quantum communication, providing unconditional security grounded in quantum mechanics. While trusted-node networks have facilitated early QKD deployment, their vulnerability to node compromise underscores the need for untrusted-node architectures. Measurement-device-independent QKD (MDI-QKD) and twin-field QKD (TF-QKD) have emerged as leading candidates, addressing security vulnerabilities and extending transmission distances. Despite the wide adoptions in various fiber scaling, no integrated implementation of these two protocols has been demonstrated to date. Here, we present a hybrid system that seamlessly integrates TF-QKD and MDI-QKD into one untrusted-node-based architecture. Utilizing an efficient phase estimation method based on asymmetric interferometers, we convert twin-field global phase tracking to relative phase calibration, allowing near continuous running of both protocols. Experiments demonstrate secure finite-size key rates for sending-or-not-sending QKD and MDI-QKD over fiber distances of 150 to 431 km. The results align with theoretical simulations and show the ability to surpass the absolute repeaterless key capacity. Our work offers an unified framework for deploying multi-protocol QKD networks, laying the foundation for adaptable and scalable quantum infrastructures that can meet a wide range of security and performance needs. 
\end{abstract}

\noindent
\section*{Introduction}
Quantum cryptography, particularly quantum key distribution (QKD) \cite{BB84} provides a way to distribute secure keys between two legitimate parties, Alice and Bob, on fiber optic networks, the secrecy of which is guaranteed by the laws of quantum mechanics \cite{Unconditional,Shor,ACM} and Vernam's one-time-pad method. Since Bennet and Brassard proposed the BB84 protocol \cite{BB84}, QKD has been studied extensively towards more established security \cite{SecRevw}, wider practicality \cite{WXB2005,Micius,drone} and quantum networking \cite{SECOQC,Wuhu,Tokyo,4600}.

Networking is crucial for large-scale QKD applications, enabling secure communication services for multi-users \cite{QaccNetwork}. In recent years, several QKD networks have been developed and demonstrated \cite{Wuhu,Tokyo,4600}, marking significant progress toward real-world implementation. These efforts have led to the successful establishment of trusted-node networks and paved the way toward the realization of a quantum internet \cite{Qinternet}. However, trusted-node networks are vulnerable to attacks that compromise the credibility of intermediary nodes, potentially paralyzing large parts of the network. If a central node becomes untrusted, the functionality of a star-topology or line-topology is compromised. It threatens the overall security of metropolitan and wide-area QKD networks based on these topologies \cite{4600,QaccNetwork}. To ensure robust security and scalability, it is imperative to transition to untrusted-node-based architectures, removing dependency on trusted nodes. 

QKD protocols with measurement-device-independent (MDI) security \cite{MDILo} are the leading candidate for upgrading quantum networks to support untrusted nodes. Compared with quantum repeaters \cite{Qrepeater} and device-independent QKD (DI-QKD) \cite{DIreview}, its maturity and ease of implementation positions these protocols as one of the most promising schemes for large-scale applications and scalable quantum networking. It effectively eliminates all detector-side channel attacks, making it highly secure and well-suited for practical deployment in untrusted-node-based networks. Unlike end-to-end QKD protocols \cite{BB84,SecRevw}, where the measurement nodes are vulnerable to security breaches, the measurement unit of these protocols can be considered untrusted, but still maintain security. 

The MDI-QKD protocol \cite{MDILo} was first introduced by Lo \textit{et al.}, leveraging the principle of virtual entanglement swapping to eliminate all detection-side vulnerabilities. The transmitters of MDI-QKD, Alice and Bob, need to prepare the same quantum states as the BB84 protocol, and the quantum states sent by both parties are transmitted through the channel to an untrusted third party, Charlie, to perform the two-photon Bell state measurement and announces the detection of valid events that Alice and Bob can use to distill secure keys. The MDI security has been rigorously proven in both infinite and finite regimes \cite{MDILo,FintMDI}. Several experiments based on decoy-state MDI-QKD protocols \cite{making,doublescan} have achieved remarkable milestones, including long-distance transmission \cite{404km,442km}, high speed \cite{HighMDI}, and on-chip systems \cite{chipMDI}. MDI-QKD has shown promise for network integration, paving the way for secure and scalable QKD networks \cite{nonstand}.

On the other hand, due to significant photon loss in quantum channel, the key rate of MDI-QKD and other end-to-end QKD protocols is fundamentally constrained by the secure key capacity of repeaterless QKD \cite{PLOB}, also known as the Pirandola-Laurenza-Ottaviani-Banchi (PLOB) bound, \( R = -\log_2(1 - \eta) \). Here, \( R \) represents the secure key rate, and \( \eta \) is the total channel transmittance between two users. To overcome this limitation, twin-field QKD (TF-QKD) \cite{TF} was proposed to achieve a key rate scaling as \( O(\sqrt{\eta}) \), surpassing the PLOB bound. In TF-QKD, both Alice and Bob independently generate and encode weak coherent pulses (WCPs) to Charlie, who then performs single-photon interference and announces the detection of valid events for Alice and Bob to distill secure keys. This process guards the protocol against eavesdropping on detection side in a manner similar to phase-encoding MDI-QKD \cite{phcodMDI}. Different from MDI-QKD, which relies on second-order optical interference and requires two-photon coincidence events to generate a raw key bit, TF-QKD utilizes first-order optical interference, allowing a raw key bit to be generated from a single-photon detection event. The full security of TF-QKD and the derivation of its secure key rate $ R $ are rigorously established through a series of theoretical advancements \cite{SNS,PM,NPP,SimpleS}, such as sending-or-not-sending QKD (SNS-QKD) \cite{SNS}. Building on these theoretics, several experimental efforts have been undertaken to demonstrate the TF-QKD's ability to surpass the repeaterless key capacity over extended fiber links \cite{WSTF,YLiu,HLiu,duband600,830km,1000km,local,comb546}. These studies highlight the potential of TF-QKD to enable longer transmission distances in QKD networks with fewer relays, and mark a critical step toward large-scale quantum communication networks.

Upgrading to untrusted-node-based QKD networks offers significant improvements in security and robustness, but it also comes with substantial costs and technical challenges that require careful evaluation. In fiber-based phase-encoding QKD networks, an important promotion would be to support both MDI-QKD and TF-QKD, ensuring a versatile, multi-protocol system that can accommodate different security needs and deployment scenarios. While TF-QKD does not inherently require asymmetric interferometers, these devices are essential in most current phase-encoding BB84 networks and MDI-QKD systems. They conveniently facilitate state preparation in the \textit{Z} and \textit{X} bases, corresponding to the Pauli operators $\sigma_Z$ and $\sigma_X$, respectively. In MDI-QKD, the encoders can be harmonized by constraining the BB84 basis choices, and the measurement unit can be upgraded for compatibility with both BB84 and MDI-QKD through the use of asymmetric interferometers \cite{nonstand}. Incorporating TF-QKD into the existing codec framework, which heavily relies on asymmetric interferometers, is therefore a critical issue for phase-encoding networks. Addressing this issue not only ensures compatibility but also lowers the barriers to deploying multi-protocol QKD systems in terms of technical demand. Despite impressive progress in extending transmission distances \cite{830km,1000km}, the integration of TF-QKD and phase-encoding MDI-QKD remains an open and pivotal task.

\begin{figure}[ht]
	\centering 
	\includegraphics[width=1.0\linewidth]{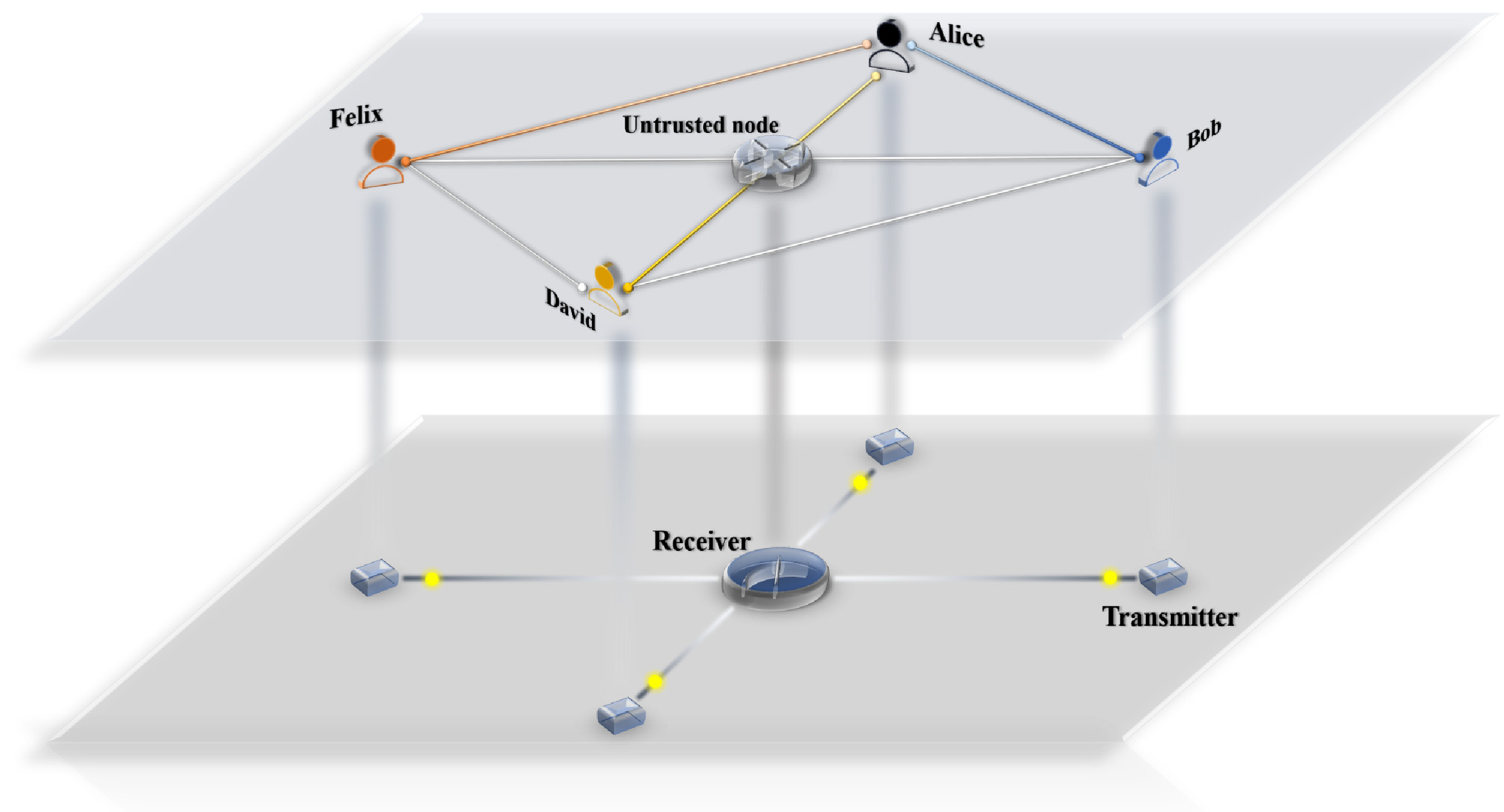}
	\caption{Schematic of the hybrid QKD network based on untrusted nodes. The red line represents the TF-QKD link, the yellow line denotes the MDI-QKD link, and the blue line indicates the multi-protocol link. The bottom layer illustrates the physical connection, where the QKD protocol is implemented between users, while the upper layer represents the users' network connections.}
	\label{fig1}
\end{figure}

Figure \ref{fig1} illustrates the schematic of a hybrid untrusted-node-based QKD network, where all users are interconnected through an untrusted central node that provides measurement services. Each user is equipped with a transmitter, and user links can dynamically select the protocol based on factors such as distance or device capabilities. For instance, Alice and David may establish an MDI-QKD link, Alice and Felix a TF-QKD link, and Alice and Bob have the flexibility to choose either protocol. The multi-protocol integration is crucial for the broad deployment of untrusted-node-based QKD networks, enabling flexible and scalable applications across diverse network conditions. In this work, we present a hybrid implementation of TF-QKD and MDI-QKD based on timebin-phase encoding. The system enables smooth switching between the two protocols and supports their continuous operation with a high transmission duty cycle, regardless of fiber length, leveraging the asymmetric interferometer structure. For a proof-of-principle demonstration, we experimentally demonstrate finite-size secure key rates for the SNS-QKD protocol with the actively-odd-parity-pairing (AOPP) method \cite{AOPP} and the MDI-QKD protocol with the double-scanning method \cite{doublescan} over a range of fiber distances. This hybrid approach provides a solution for multi-protocol compatibility of untrusted-node-based QKD networks.

\section*{Results}
\subsection*{Experimental design}
\begin{figure}[ht]
	\centering 
	\includegraphics[width=1.0\linewidth]{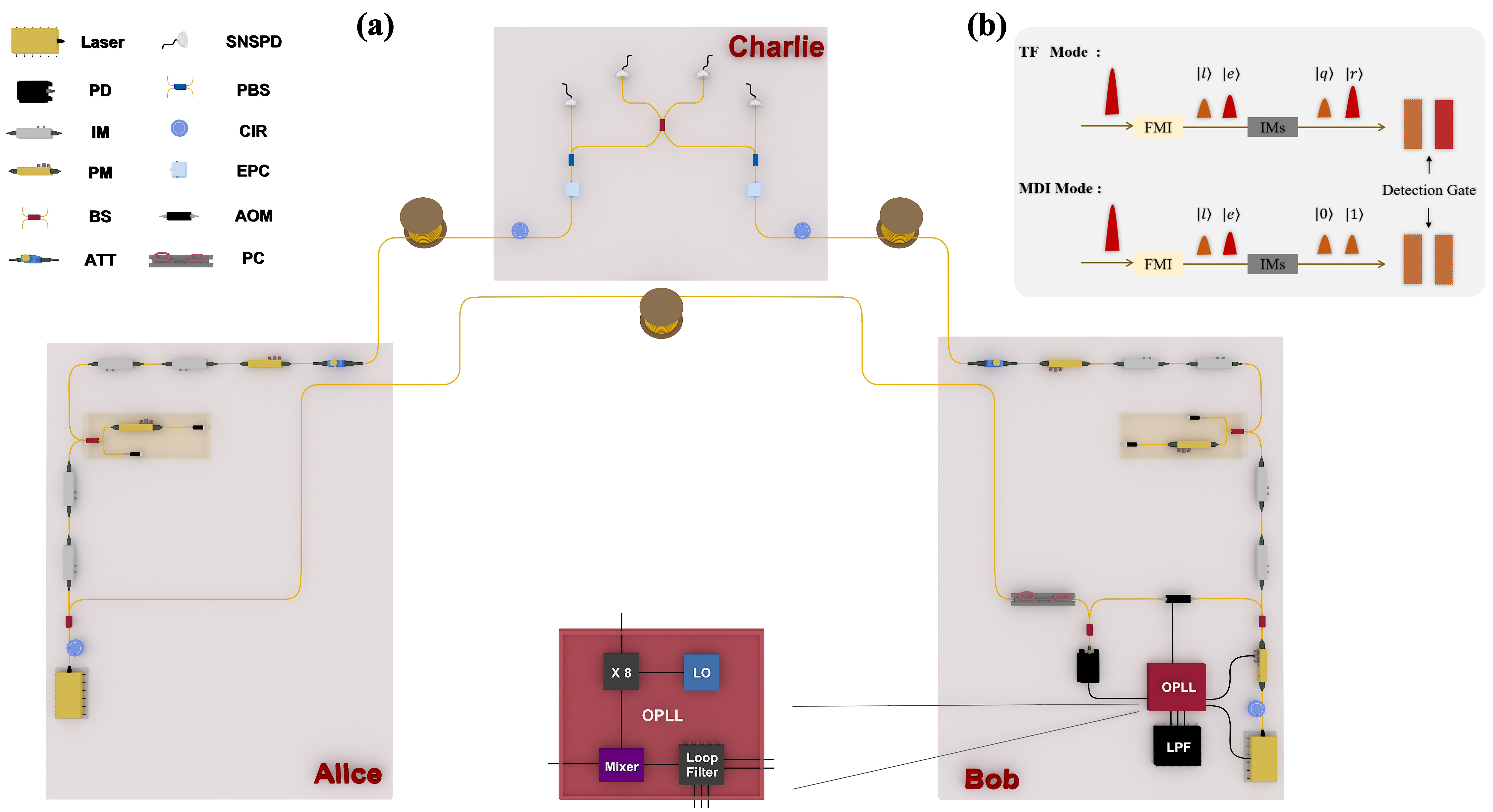}
	\caption{Experimental setup. SNSPD: superconducting nanowire single-photon detector; PD: photodetector; PBS: polarization beam splitter; IM: intensity modulator; CIR: circulator; PM: phase modulator; EPC: electronic polarization controller; BS: beam splitter; AOM: acousto-optic modulator; ATT: attenuator; PC: polarization controller; LO: local oscillator; LPF: low-pass filter; FMI: Faraday-Michelson interferometer. }
	\label{fig2}
\end{figure}

A conceptual schematic of the experimental setup is provided in Fig. \ref{fig2}a. On the source side, two narrow-linewidth semiconductor continuous-wave lasers (Rio PLANEX) with a central wavelength of 1549.32 nm and a linewidth of approximately 2 kHz are frequency-locked to each other using a heterodyne optical phase-locked loop (OPLL). Each laser is capable of maintaining optical frequency stability within 5 MHz over six hours. Alice's continuous light is divided by a 50:50 beam splitter (BS), while Bob’s continuous light, after passing through a phase modulator (PM), undergoes a similar division by a 50:50 BS. One beam from each side is used for quantum-state encoding, while another beam from Alice is transmitted through the servo channel (0.18 dB/km) and coupled with the corresponding beam from Bob to generate a beat note. A polarization controller (PC) on Bob side aligns the polarization of the beams involved in the beat frequency, maximizing the magnitude of the beat signal. An acousto-optic modulator (AOM) is positioned after the BS to introduce a precise optical frequency shift of 300.7 MHz, which also serves as the reference frequency. By locking the beat note to this signal, the OPLL and the PM0 on Bob side achieve precise optical phase locking. The evaluation of OPLL is discussed in detail in sec. \uppercase\expandafter{\romannumeral 3}. 

The locked coding beams at both Alice's and Bob's sides are processed through two cascaded intensity modulators (IMs), which chop the light into pulse sequences with a 50 MHz repetition rate and a temporal width of 1.5 ns, achieving an extinction ratio exceeding 30 dB. These pulses are directed into Faraday-Michelson interferometers (FMIs) \cite{YPChen,442km}, which split each incident pulse into two time-bin pulses separated by a 10 ns interval: the earlier time-bin pulse $ \lvert e \rangle $ and the later time-bin pulse $ \lvert l \rangle $. A circulator placed before each FMI filters out any laser light reflected backward. For operations in MDI-QKD mode, IM3 and IM4 handle basis selection and decoy-state modulation. Phase coding in the \textit{X} basis is performed by adding a relative phase \{$ 0 $, $ \pi $\} to the $ \lvert l \rangle $ pulse via PM1, while time-bin coding in the \textit{Z} basis is achieved by chopping either the earlier or later time-bin pulse, corresponding to a bit value of 0 or 1, respectively. These IMs effectively suppress ambient noise, ensuring a high extinction ratio for both the vacuum state and the eigenstates of the \textit{Z} basis, resulting in an inherent error rate of just 0.1\% for \textit{Z} basis. Finally, PM2 is used to randomize the phase of both time-bin pulses, mitigating potential attacks on the light sources.

For operations in TF-QKD mode, IM3 and IM4 modulate and encode the later time-bin pulse $ \lvert l \rangle $ into the quantum pulse $ \lvert q \rangle $ with random intensities while retaining the earlier time-bin pulse $ \lvert e \rangle $ as the reference pulse $ \lvert r \rangle $ for phase calibration. The majority of duty periods in TF-QKD mode are allocated for transmission, with a smaller portion reserved for compensation. During transmission, phase encoding \{$ 0 $, $ \pi $\}, representing 0 or 1 bit, respectively, is applied to $ \lvert q \rangle $ by PM1, along with a random phase slice of $ 2\pi i/16, \ i \in \{1, \ldots, 16\} $. Simultaneously, PM2 alternately scans phases $ 0 $ and $ \pi/2 $ solely on $ \lvert r \rangle $, referred to as two-phase scanning, to acquire global phase drifts across the quantum channel. A detailed description of the phase estimation method is provided in Sec. \uppercase\expandafter{\romannumeral 3}. During compensation, the two-phase scanning procedure is applied simultaneously to both $ \lvert q \rangle $ and $ \lvert r \rangle $, with PM1 ceasing its encoding functions, while IM3 and IM4 modulate both pulses into the reference intensity. The same procedure is used for phase calibration in MDI-QKD mode, yielding a interference visibility of approximately 99.5\% in the \textit{X} basis. An optical delay (OD) is inserted on Alice side after the IMs to fine-tune the photon arrival time with picosecond precision. Finally, an electrically tunable optical attenuator (EVOA) attenuates the pulses to the single-photon level before they are transmitted to Charlie through ultralow-loss fibers (0.17 dB/km).

On the measurement side, the incident pulses from Alice and Bob are first routed through a polarization-maintaining module comprising an electronic polarization controller (EPC) and a polarization beam splitter (PBS) to ensure consistent polarization of the outgoing pulse pairs. These pairs are then coupled at a BS for interference and subsequently detected by superconducting nanowire single-photon detectors (SNSPDs) from PHOTEC (model P-SPDNS). Both detectors, D0 and D1, feature a detection efficiency of 68.5\% and a dark count rate of 10 Hz. After accounting for the insertion loss of all components, the overall detection efficiency is 64\%, and the dark count rate per pulse is approximately $ 1.5 \times 10^{-8} $ when gated with a 1.7 ns time window. In TF-QKD mode, single-photon detection events are recorded by a time-to-digital converter (TDC), with distinct time windows assigned for quantum and reference pulses to facilitate post-processing. In MDI-QKD mode, detection events from D0 and D1 are recorded to calculate coincidence events, enabling the Bell state measurement. By switching the encoding modes of Alice and Bob, the hybrid system smoothly transitions between TF-QKD and MDI-QKD.

\subsection*{Protocols Description}
We follow the SNS-QKD protocol with AOPP method \cite{AOPP} and the MDI-QKD protocol with double-scanning method \cite{doublescan} in this experiment. For TF-QKD mode, the \textit{X} basis is encoded in phase of WCPs, and \textit{Z} basis is encoded in intensity \{$ \mu $, $ \nu $, $ \omega $, $ o $\}, where $ \mu $ and $ o $ denote the signal and vacuum intensity, respectively, and $ \nu $ and $ \omega $ are the decoy intensities. Alice (Bob) randomly decides whether the \textit{i}th time window is a decoy window or a signal window. For a signal window, Alice (Bob) randomly chooses $ \mu $ ($ o $) with probabilities $ \epsilon $ ($ 1-\epsilon $), indicating a sending (not-sending) choice in SNS-QKD protocol. Alice (Bob) actually prepares a phase-randomized WCP with intensity $ \mu $. For a decoy window, Alice (Bob) randomly chooses sources from $ \nu $, $ \omega $ and $ o $, and prepares a phase-randomized WCP. The prepared pulse pairs are send to Charlie to perform interferometric measurements. For MDI-QKD mode, the \textit{X} basis is encoded in the relative phase of earlier and latter pulse, and \textit{Z} basis is encoded in its time-bins. Signal intensity $ \nu $ is only prepared in \textit{Z} basis while decoy intensities $ \nu $ and $ \omega $ are in \textit{X} basis. Both Alice and Bob do not choose any bases for vacuum states $ o $. The prepared quantum states are similarly send to Charlie to perform the $ \lvert\psi^{-}\rangle $ projection measurement. All intensities and probabilities are optimized by the differential evolutionary algorithm. As a proof-of-principle demonstration, we encode the WCPs into 16 phase slcies between $ 0 $ and $ 2\pi $, and generate an 2000-bit pseudo-random pattern, which is preloaded into an arbitrary waveform generator (AWG) to drive the modulators repeately. Tab. \ref{tab1} lists the system parameters adopted in this experiment.

\begin{table}[htp]
	\centering
	\caption{List of system parameters. Here $ \eta_{d} $ is detection efficiency, $ f $ is the error correction inefficiency, $ e_{PS} $ is the misalignment error with phase slice, $ e_{X} $ is the misalignment error of \textit{X} basis in MDI-QKD, $ e_{Z} $ is the misalignment error of \textit{Z} basis, $ \varepsilon $ is the the failure probability, and $ p_{d} $ is the dark count rate.}
	\begin{tabular}{p{1.2cm}p{1.2cm}p{1.2cm}p{1.2cm}p{1.2cm}p{1.2cm}p{2.0cm}}
		\hline
		\hline
		$ \eta_{d} $ & $ f $ & $ e_{PS} $ & $ e_{X} $ & $ e_{Z} $ & $ \varepsilon $ & $ p_{d} $ \\
		\hline
		$ 64\% $ & $ 1.1 $ & $ 4.10\% $ & $ 0.97\% $ & $ 0.11\% $ & $ 10^{-10} $ & $ 1.5\times10^{-8} $  \\	
		\hline
		\hline
	\end{tabular}
	\label{tab1}
\end{table}

\subsection*{Experimental Results}
\begin{figure}[htp]
	\centering
	\includegraphics[width=1.0\linewidth]{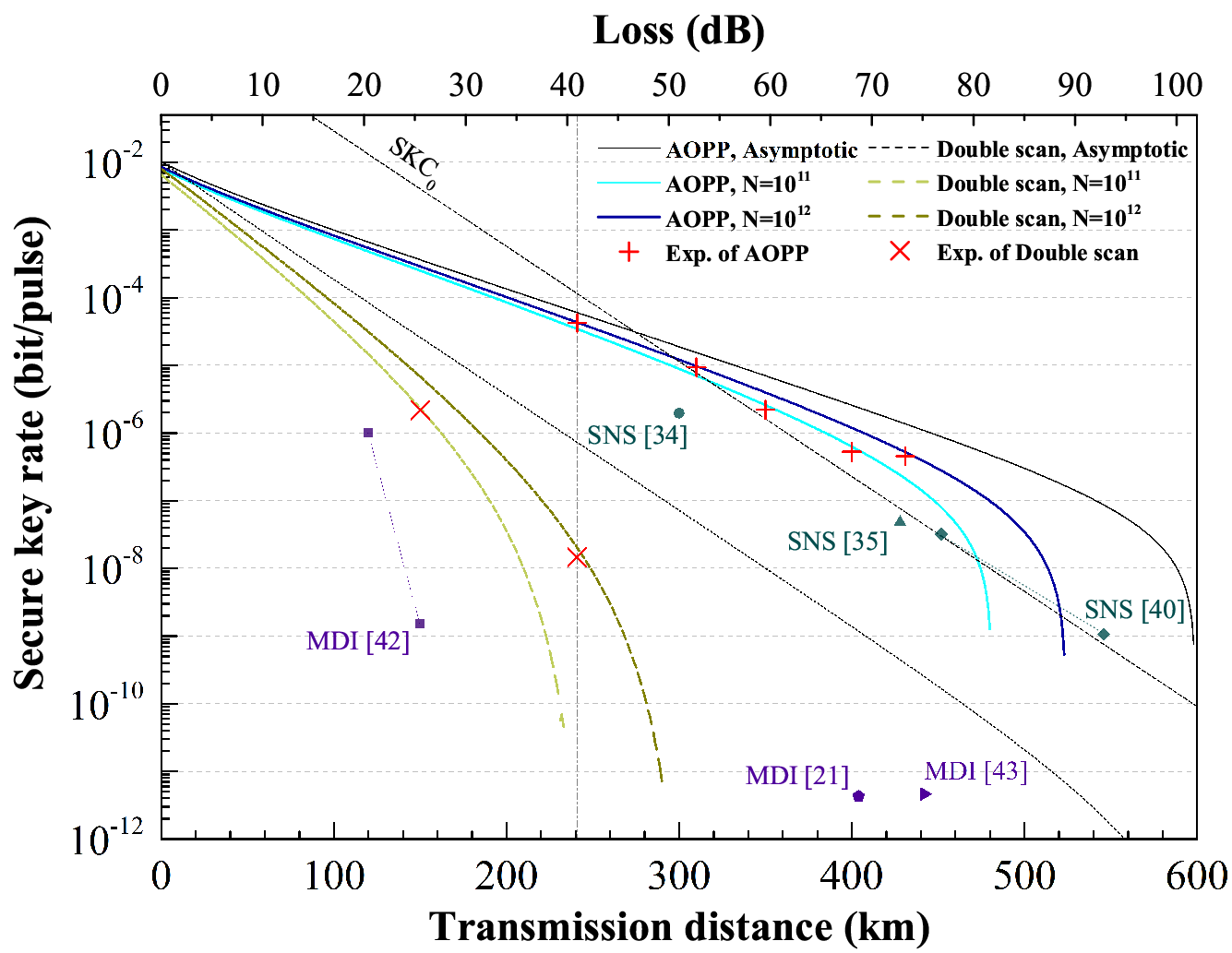}
	\caption{Simulation and experimental results with $ N=10^{11} $ and $ N=10^{12} $. The SKC$_{0}$ is the absolute repeaterless bound \cite{PLOB} of the channel loss. Key rates per pulse of previous MDI-QKD experiments and TF-QKD experiments are included for comparison based on their transmission distances.}
	\label{fig5}
\end{figure}

The SNS-QKD protocol and MDI-QKD protocol are performed over a range of different channel distance. The secure key rates calculated in the finite-size regime from the experiment and theoretical simulations shown in Fig. \ref{fig5}, together with the absolute PLOB bound \cite{PLOB} which is calculated as $ -log_{2}(1-\eta) $ with 100\% detection efficiency. The asymptotic case refers to the analysis of QKD when the total number of pulses \textit{N} tends to infinity. The asymptotic lines present a performance potential of the hybrid system in both protocols. In practical implementations, the finite-size analysis provides the security against statistical fluctuations for the final secure keys, bounded by the Chernoff bound \cite{doublescan,AOPP}. 

We therefore conduct experiments at total distances from 150 km to 431 km between Alice and Bob in a symmetric fiber setup. Positive secure keys are generated for all these distances in finite-size regime, i.e., $ N=10^{11} $ and $ N=10^{12} $ ($ 5\times10^{7} $ and $ 5\times10^{8} $ pattern repetition), and the detailed results can be found in Supplementary text. The experimental secure key rates of SNS-QKD scales as $ O(\sqrt{\eta}) $ and are in agreement with the theory. From 310 km to 431 km, the secure key rates break the absolute PLOB bound in both pulse sizes. The experimental secure key rates of MDI-QKD are measured at 150 km and 241 km with either pulse sizes, which hold agreements with the theoretical simulations. At 241 km fiber distance, the hybrid system sends $ N=10^{9} $ repeated patterns continuously. For the first $ N=5\times10^{8} $, the system runs the SNS-QKD protocol and encodes the WCPs, while for the remaining repeated patterns, the system runs the MDI-QKD protocol and encodes the four kinds quantum states. In both protocols, Charlie simultaneously records the single-photon clicks of each detector window. And the global phase recovery and two-photon coincidence events are calculated by post-processing before sifting. We adopt the AOPP method to improve the key rates of SNS-QKD, which is a kind of two-way classical communication. 

By utilizing the filter-enhanced two-phase scan, phase compensation is carried out for 200 $ \mu $s every 10 seconds in both protocols at each distance, achieving an duty cycle of near 99\% after accounting clock synchronizing and polarization compensation. The obtained secure key rates of TF-QKD in key bit per second (per pulse) are 2105.59 bps, 463.52 bps, 108.91 bps, 26.16 bps and 22.60 bps at 241 km, 310 km, 351 km, 400 km and 431 km fiber distance, respectively. And the obtained secure key rates of MDI-QKD in key bit per second are 107.61 bps and 0.72 bps at 150 km and 241 km, respectively. The experimental results demonstrate the feasibility of the hybrid implementation for QKD networks with untrusted-nodes. With a larger pulse number and a lower dark count rate, the current transmission distance can further achieve a distance similar to the Ref. \cite{comb546} and \cite{442km}.

\section*{Discussion}
In conclusion, we have demonstrated a hybrid implementation of TF-QKD and MDI-QKD within a single timebin-phase encoding system, achieving seamless switching between phase-encoding MDI-QKD and TF-QKD. To integrate and improve the applicability of these two protocols, we have developed an efficient global phase estimation scheme by utilizing the preserved asymmetric interferometers, simplifying the global phase tracking of twin fields to phase calibration comparable to that of MDI-QKD. This approach allows both protocols to operate continuously for durations on the order of ten seconds and achieve a transmission duty cycle exceeding 99\%, independent of fiber length. By adopting the optical phase locking, we experimentally demonstrate the SNS-QKD protocol with the AOPP method and the MDI-QKD protocol with the double-scanning method over fiber distances ranging from 150 km to 431 km. The hybrid system achieves positive key rates for both protocols in the finite-size regime and surpasses the absolute repeaterless key capacity across a range of channel distances. The experimental results demonstrate the feasibility and high efficiency of the proposed scheme. Our implementation sets a new benchmark for establishing a versatile and unified framework that enables the integration of multiple protocols and ensuring long-term compatibility for future advancements in QKD networks.

As discussed, frequency locking and phase tracking over deployed fibers remain critical issues for TF-QKD. The hybrid implementation can accommodate alternative frequency locking methods by simply altering the laser sources, such as ultra-stable lasers with Pound-Drever-Hall locking \cite{YLiu}, optical injection locking \cite{HLiu}, and the recently proposed wavelength locking with a local reference frequency \cite{local}. Additionally, our scheme demonstrates the potential for integration with dual-band phase stabilization \cite{duband600,1000km,comb546}, which has been successfully applied in high-loss and long-distance systems \cite{1000km}. Optimizing the path of dual-band pulses within the encoder, such that the quantum and reference light paths mimic the short and long arms of an interferometer, could theoretically reduce dual-band scanning time and further improve the system's duty cycle. Incorporating on-chip Mach-Zehnder interferometers into the hybrid implementation offers further anticipated performance enhancement. This system also supports the recently proposed MP-QKD protocol \cite{MP}, which allows reduced reliance on phase locking and global phase tracking, and can provide either performance comparable to MDI-QKD (with a fixed matching interval of 2) or a rate-loss scaling similar to TF-QKD (with a larger matching interval, such as 2000 in Ref. \cite{MPexp}), depending on user requirements. Moreover, the hybrid system is compatible with upgrading existing phase-encoding BB84 QKD networks to untrusted-node QKD networks at minimal hardware cost. For instance, by integrating the receiver scheme in Ref. \cite{nonstand}, the architecture has the ability to support BB84 QKD, MDI-QKD, and TF-QKD while retaining the original codecs in existing QKD networks. This underscores the immense potential of integration in trusted-node-based and untrusted-node-based QKD networks, paving the way towards the next generation of flexible and multifunctional quantum communication infrastructures.


\section*{Materials and Methods}
\subsection*{Optical Phase-Locked Loop}
The schematic of the OPLL is depicted in the inset of Fig. \ref{fig2}a. Inside the OPLL, a local oscillator (LO) clock at 37.5875 MHz is generated by the high-precision onboard clock of the field-programmable gate array (Xilinx ZCU111 development board). This signal is then frequency-multiplied by 8 using an electronic phase-locked loop (PLL) and amplified by a radio-frequency (RF) amplifier, producing the 300.7 MHz RF signal that drives the AOM and serves as the reference clock for the mixer in the OPLL system. The beat note, converted through a 12 GHz bandwidth photodetector (Optilab PR-12-B-M), along with the reference clock, is fed into a loop filter, providing feedback for the proportional-integral-derivative (PID) algorithm ($g_p = -6$ dB, $f_I = 16$ kHz, $f_D = 250$ kHz). The PID algorithm calculates the necessary adjustments to the control signal, performing fast wavelength modulation of Bob’s laser with the direct current signal in the range of $ \pm4 $ V, ensuring a small and stable frequency difference between the two lasers. The residual phase noise is then processed by the PM0 on Bob side, which is controlled by a low-pass filter. The measured relative Allan deviation of the beat note in the time domain, ranging from \( 10^{-11} \) to \( 10^{-13} \), demonstrates excellent frequency stability. The measurement results as well as the beat note spectra are given in supplementary text. In the current setup, the residual mean-square phase error associated with the OPLL is about \( 5.6 \times 10^{-3} \ \mathrm{rad^2} \).

\subsection*{Efficient Phase Estimation based on Asymmetric Interferometers}
In this section, we showcase the efficient post phase-estimation method by using the FMIs. As illustrated in Fig. \ref{fig2}b, after passing through the FMI, the optical pulses are split into two time-bin pulses: the earlier bin pulse $ \lvert e \rangle $ and the latter bin pulse $ \lvert l \rangle $. The earlier pulses are modulated into reference pulses $ \lvert r \rangle $ using IMs, while the latter pulses are modulated into quantum pulses $ \lvert q \rangle $. Then, the global phase of reference pulses $ \lvert r \rangle $ when arriving at the beam splitter (BS) on Charlie side can be written as
\begin{align}\label{Phase_ref}
	\phi_{ref}=\phi_{0}+\varphi_{short}+\phi_{channel},
\end{align}
where $ \phi_{0} $ denotes the initial phase of lasers plus the optical path before input into the interferometers, $ \varphi_{short} $ denotes the phase applied by the short arm of interferometers, and $ \phi_{channel} $ denotes the phase applied by the quantum channel link after outgoing the interferometers. Similarly, the global phase of quantum pulses $ \lvert q \rangle $ when arriving at the BS on Charlie side is
\begin{align}\label{Phase_qut}
	\phi_{q}=\phi_{0}+\varphi_{long}+\phi_{channel},
\end{align}
where $ \varphi_{long} $ denotes the phase applied by the long arm of interferometers. In our scheme, since the reference pulses $ \lvert r \rangle $ and quantum pulses $ \lvert q \rangle $ are emitted from the same laser on either the Alice or Bob side, and travel through nearly identical channels with consistent polarization over a period of transmission time, the initial laser phase and the phase of quantum channel can be considered nearly equal \cite{AMZI,FMI}. The path difference between the long and short arms of FMI allows us to adjust the phase applied by the PM inserted in either arm, thereby equalizing $ \phi_{ref} $ and $ \phi_{q} $ within a certain period. 

On the detection side, the D0 and D1 detectors operate with two separate time-bin windows to distinguish between reference and quantum pulses. The interference results of the reference pulses from Alice and Bob are projected onto the earlier time-bin windows of the D0 and D1 detectors, enabling the estimation of global phase drifts across the quantum channel. Over extended periods, however, the phase differences between the reference and quantum pulses are primarily attributed to fluctuations in the optical paths of the two arms of the interferometers. Theoretically, this problem can be solved by adopting phase calibrations similar to phase-encoding BB84 QKD or MDI-QKD utilizing asymmetric interferometers.

In our experiment, to compensate for the phase difference between pulse $ \lvert r \rangle $ and $ \lvert q \rangle $, Alice and Bob first modulate their both pulses into reference intensity. Either Alice or Bob performs the filter-enhanced two-phase scan using a PM located after the FMI, covering the arrival times of both $ \lvert r \rangle $ and $ \lvert q \rangle $. During the scan, the PM inside the FMI sweeps through $ I = 16 $ phase slices, recording the corresponding scanning statistics for both $ \lvert r \rangle $ and $ \lvert q \rangle $. Subsequently, it selects and applies the phase slice that minimizes the counting difference between $ \lvert r \rangle $ and $ \lvert q \rangle $, calibrating the phase drifts of the two arms of the interferometer. During transmission, Alice and Bob modulate $ \lvert q \rangle $ into signal, decoy, or vacuum intensities, applying phase randomization and encoding solely to $ \lvert q \rangle $. Meanwhile, the PM positioned after the FMI continues to load the scanning signal and carries out the filter-enhanced two-phase scan exclusively on the time sequence of the $ \lvert r \rangle $ pulses. All perturbations in the fiber channel affect $ \lvert r \rangle $ and $ \lvert q \rangle $ equally, enabling us to estimate $ \phi_{q} $ with $ \phi_{ref} $ and minimizing interruptions of quantum pulse transmission. 

\begin{figure}[ht]
	\centering
	\includegraphics[width=1.0\linewidth]{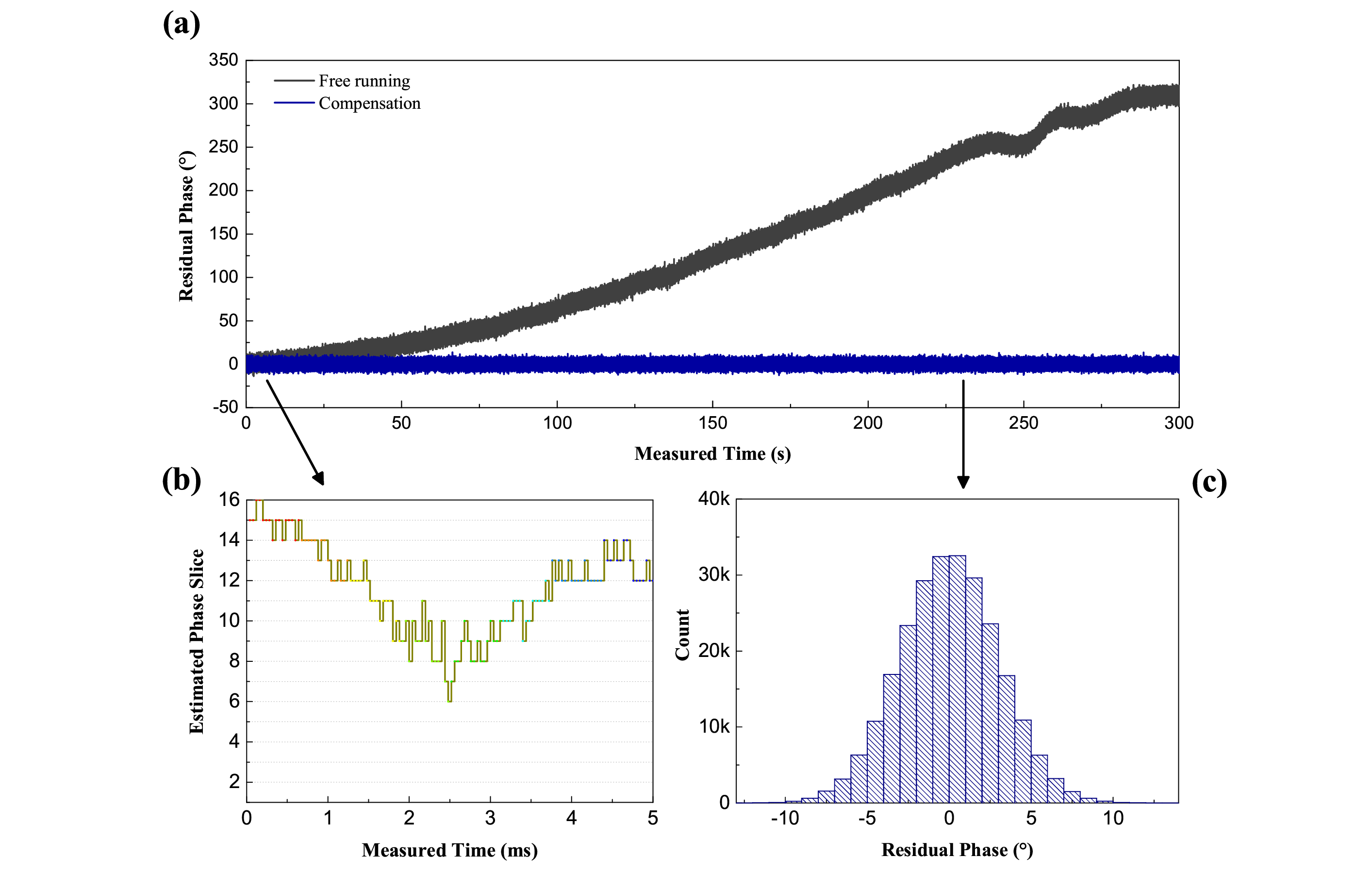}
	\caption{(a) Monitoring $ \varDelta\phi_{error} $ over time with and without compensation at a 300 km fiber distance; (b) Phase slice variations of pulse $ \lvert r \rangle $ and $ \lvert q \rangle $ within the first 5 ms; (c) Histogram of $ \varDelta\phi_{error} $ with compensation applied every 10 seconds.}
	\label{fig3}
\end{figure}

Drifts in the long and short arms of the FMI can disrupt this equivalence; thus, it remains necessary to periodically repeat the above steps to compensate for phase shifts between pulse $ \lvert r \rangle $ and $ \lvert q \rangle $. This process ensures minimal QBER and reduces residual estimation error, which is defined as
\begin{align}\label{residualerror}
	\varDelta\phi_{error}=|\phi_{q}-\phi_{ref}|.
\end{align}
We test this method over a range of fiber distance with an execution time of 40 $ \mu $s for the two-phase scan. Fig. \ref{fig3}(a) shows the monitoring results of $ \varDelta\phi_{error} $ with and without compensation at 300 km. It demonstrate that $ \varDelta\phi_{error} $ can consistently maintain a stable phase slice for at least 10 seconds across long fiber distances. Fig. \ref{fig3}(b) shows the phase slice variations during the first 5 ms of global phase estimation from both the reference pulses $ \lvert r \rangle $ and quantum pulses $ \lvert q \rangle $, modulated with the same reference intensity. The overlap of these two stair curves highlights the effectiveness of this method. For a 10-second compensation period, the histogram of $ \varDelta\phi_{error} $ exhibits a standard deviation of 3.98°, as shown in Fig. \ref{fig3}(c). In this configuration, the hybrid system can run continuously for several seconds in TF-QKD mode, requiring only 200 $ \mu $s of five round scanning to compensate for the arm length fluctuations in the FMI, significantly enhancing the system’s duty cycle.

\subsection*{Filter-enhanced two-phase scan}
In previous TF-QKD experiments \cite{WSTF,YLiu,HLiu,duband600,830km,1000km,local,comb546}, a time gap of several hundred nanoseconds was typically introduced between strong reference frames and weak quantum pulses to allow SNSPDs to recover. In our scheme, the close temporal proximity of quantum and reference pulses negatively affects the signal-to-noise ratio (SNR), leading to increased misalignment errors. To address this, we integrate a least mean square (LMS) adaptive filter with the two-phase scan, enabling accurate estimation even at reduced reference intensity and effectively mitigating the impact of double Rayleigh scattering noise on misalignment errors. The noisy count acquired in two-phase scan can be expressed as \( N(M) = \hat{N}(\hat{M}) + \tilde{\epsilon} + \tilde{\delta} \), where \( \hat{N}(\hat{M}) \) is the expected count, \( \tilde{\epsilon} \) is Gaussian-distributed additive noise, and \( \tilde{\delta} \) accounts for systematic errors caused by shot-noise and overshoot amplitudes. After sweeping through $ I $ phase slices while conducting $ I $ times two-phase scan, we obtain two sequences of input $ \textbf{x}_{r}(n)=[N_{0}^{r}, N_{1}^{r}, M_{0}^{r}, M_{1}^{r}]^{T} $ and $ \textbf{x}_{q}(n)=[N_{0}^{q}, N_{1}^{q}, M_{0}^{q}, M_{1}^{q}]^{T} $, $ n\in1,...I $ where $ N_{0}^{r(l)} $ and $ N_{1}^{r(l)} $ denote the recording detector D0's count of pulses $ \lvert r \rangle $ ( $ \lvert q \rangle $ ) when loading phase $ 0 $ and $ \pi/2 $ within the first 20 $ \mu $s and the second 20 $ \mu $s, respectively, and $ M_{0}^{r(l)} $ and $ M_{1}^{r(l)} $ denote detector D1's count of pulses $ \lvert r \rangle $ ( $ \lvert q \rangle $ ) when loading phase $ 0 $ and $ \pi/2 $ within the first 20 $ \mu $s and the second 20 $ \mu $s, respectively. The filter is controlled by a set of 10 weights, denoted as $ \textbf{w}=[w_{1}, ..., w_{10}]^{T} $. By convolving the input signal with $ \textbf{w} $, we get the expected output sequences of the filter 
\begin{align}\label{convolv}
	\textbf{y}_{r}(n)=\textbf{w}^{T}\ast \textbf{x}_{r}(n) =
	[\hat{N}_{0}^{r}, \hat{N}_{1}^{r}, \hat{M}_{0}^{r}, \hat{M}_{1}^{r}]^{T}, \\
	\textbf{y}_{q}(n)=\textbf{w}^{T}\ast \textbf{x}_{q}(n) = [\hat{N}_{0}^{q}, \hat{N}_{1}^{q}, \hat{M}_{0}^{q}, \hat{M}_{1}^{q}]^{T}.
\end{align}
Then, $ \textbf{y}_{r}(n) $ and $ \textbf{y}_{q}(n) $ sequences are calculated by the two-phase scan method given in the supplementary text. Denote the output of two-phase scan algorithm as $ \textbf{s}_{r(q)}(n)=2PS(\textbf{y}_{r(q)}(n)) \in 1,...I $, we can define the desired signal $ \hat{\textbf{d}}(n) $ as 
\begin{align}\label{hatd(n)}
	\hat{\textbf{d}}(n)=cos[\dfrac{\pi \cdot |\textbf{s}_{r}(n)-\textbf{s}_{q}(n)|}{I}]^{2}.
\end{align}
And the reference signal is $ \textbf{d}(n)=cos[\pi(|n-\mathit{K}|)/I]^{2} $, in which $ \mathit{K} $ is the phase slice order corresponding to the minimum value of $ |\textbf{x}_{r}(n)-\textbf{x}_{q}(n)| $. It is a linear constraint on the differences between the phase slices of pulse $ \lvert r \rangle $ and $ \lvert q \rangle $ considering no noise. Then, the mean square error is given as $ J=E[\textbf{e}^{2}(n)] $, where $ \textbf{e}(n)=\textbf{d}(n)-\hat{\textbf{d}}(n) $ is the cost function. Utilizing the steepest descent algorithm, the updated weight vector $ \textbf{w}_{upd} $ is adjusted along the negative gradient direction during every compensation, given as
\begin{align}\label{w_upd}
	\textbf{w}_{upd}=\textbf{w}+ - \lambda\dfrac{\partial\textbf{e}^{2}(n)}{\partial\textbf{w}},
\end{align}
where $ \lambda=7.5\times10^{-3} $ is the step-size. Through adaptive feedback cancellation, the LMS filter effectively reduces the effect of statistical fluctuation of fewer counts on the accuracy of estimation, and mitigates modulation deviation of PMs and intensity fluctuations from laser source. And the initialization of LMS filter is conducted with the data pre-acquired from an attenuation-simulated channel by traversing the entire voltage range of the PM to extract the expected output for precise calibration.

\begin{figure}[ht]
	\centering
	\includegraphics[width=1.0\linewidth]{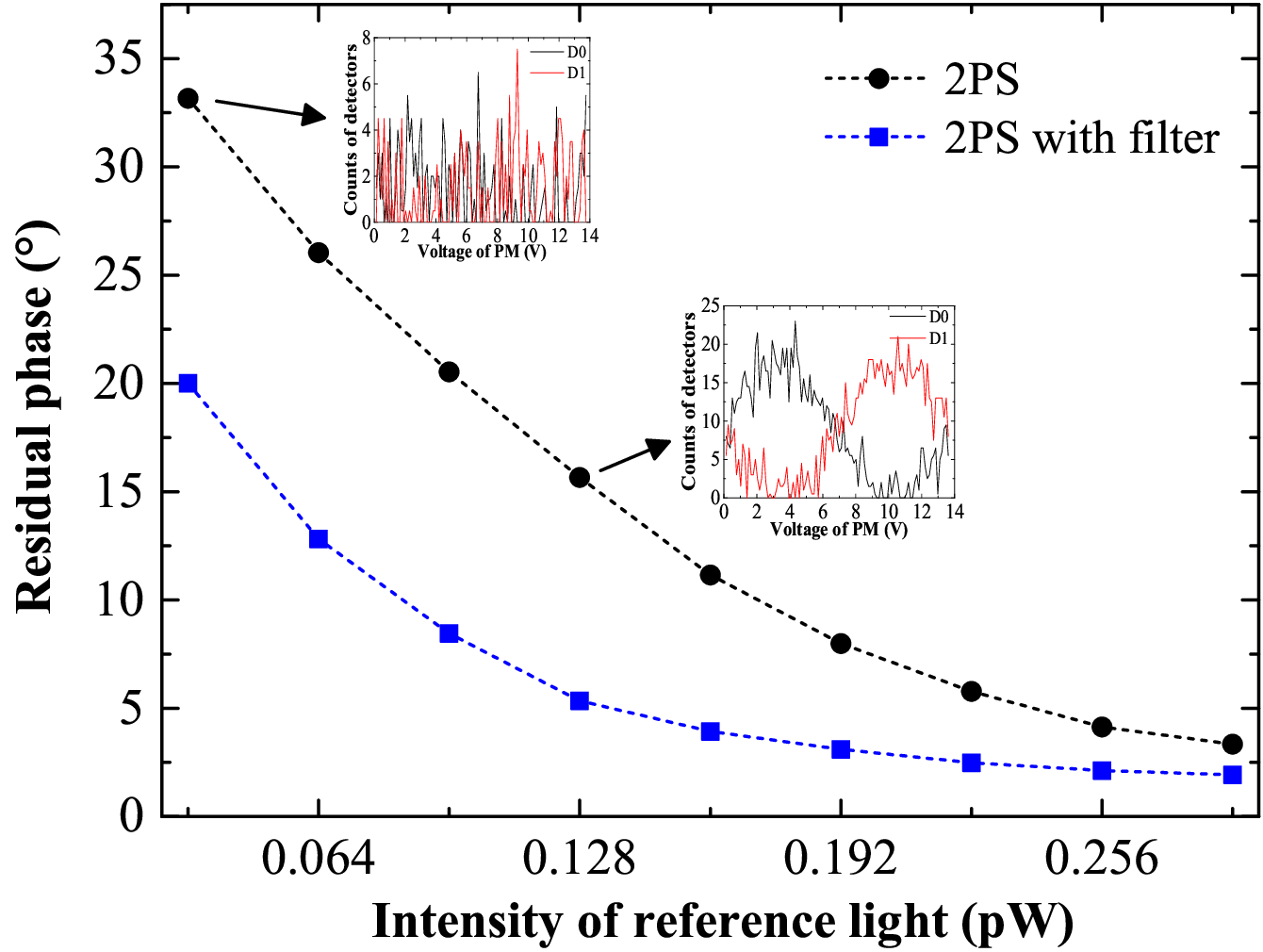}
	\caption{MAE of two-phase scan (2PS) with and without the LMS filter with respect to reference light intensity.}
	\label{fig4}
\end{figure}

We examine the mean absolute error (MAE) between the global phase estimates with and without the filter-enhanced two-phase scan method under varying reference light intensities, as shown in Fig. \ref{fig4}. Considering the trade-off between MAE and Rayleigh scattering noise, the reference intensity is set as about 0.16 pW through a 400 km fiber, corresponding to the total SNSPD counts of 1.0 MHz. The two insets of Fig. \ref{fig4} show the shape of interference curves acquired at the total counts of 0.5 MHz and 2.5 MHz, respectively. The results demonstrate that the proposed method can effectively reduce reference light intensity while maintaining estimation accuracy.

\clearpage 

%

\begin{thebibliography}{1}
	
\bibitem{BB84} C. H. Bennett and G. Brassard, Quantum cryptography: public key distribution and coin tossing. In Conference on Computers, Systems and Signal Processing (1984), pp. 175–179.
	
\bibitem{Unconditional} H. K. Lo, and H. F. Chau, Unconditional Security of Quantum Key Distribution over Arbitrarily Long Distances. Science \textbf{283}, 2050-2056 (1999).

\bibitem{Shor} P. W. Shor, and J. Preskill, Simple proof of security of the BB84 quantum key distribution protocol. Phys. Rev. Lett. \textbf{85}, 441 (2000).

\bibitem{ACM} D. Mayers, Unconditional security in quantum cryptography. J. ACM. \textbf{48}, 351 (2001).

\bibitem{SecRevw} V. Scarani, H. Bechmann-Pasquinucci, N. J. Cerf, \textit{et al.}, The security of practical quantum key distribution. Rev. Mod. Phys. \textbf{81}, 1301 (2009).

\bibitem{WXB2005} X. B. Wang, Beating the photon-number-splitting attack in practical quantum cryptography. Phys. Rev. Lett. \textbf{94}, 230503 (2005).

\bibitem{Micius} S.-K. Liao, W.-Q. Cai, W.-Y. Liu, \textit{et al.}, Satellite-to-ground quantum key distribution. Nature (London) \textbf{549}, 43 (2017).

\bibitem{drone} X.-H. Tian, R. Yang, H.-Y. Liu, \textit{et al.}, Experimental demonstration of drone-based quantum key distribution. Phys. Rev. Lett. \textbf{133}, 200801 (2024).	

\bibitem{SECOQC} M. Peev, C. Pacher, R. Alléaume, \textit{et al.}, The SECOQC quantum key distribution network in Vienna. New J. Phys. \textbf{11}, 075001 (2009).

\bibitem{Wuhu} F. X. Xu, W. Chen, S. Wang. \textit{et al.}, Field experiment on a robust hierarchical metropolitan quantum cryptography network. Chin. Sci. Bull. \textbf{54}, 2991–2997 (2009).

\bibitem{Tokyo} M. Sasaki, M. Fujiwara, H. Ishizuka, \textit{et al.}, Field test of quantum key distribution in the Tokyo QKD Network. Opt. Express \textbf{19}, 10387-10409 (2011). 

\bibitem{4600} Y. A. Chen, Q. Zhang, T. Y. Chen, \textit{et al.}, An integrated space-to-ground quantum communication network over 4,600 kilometres. Nature \textbf{589}, 214–219 (2021).

\bibitem{QaccNetwork} B. Fröhlich, J. F. Dynes, M. Lucamarini, \textit{et al.}, A quantum access network. Nature \textbf{501}, 69–72 (2013).

\bibitem{Qinternet} S. Wehner, D. Elkouss, and R. Hanson, Quantum internet: a vision for the road ahead. Science \textbf{362}, eaam9288 (2018).

\bibitem{MDILo} H. K. Lo, M. Curty, and B. Qi, Measurement-device-independent quantum key distribution. Phys. Rev. Lett. \textbf{108}, 130503 (2012).

\bibitem{Qrepeater} N. Sangouard, C. Simon, H. De Riedmatten, \textit{et al.}, Quantum repeaters based on atomic ensembles and linear optics. Rev. Mod. Phys. \textbf{83}, 33–80 (2011).

\bibitem{DIreview} V. Zapatero, T. van Leent, R. Arnon-Friedman, \textit{et al.}, Advances in device-independent quantum key distribution. npj Quantum Inf. \textbf{9}, 10 (2023).

\bibitem{FintMDI} M. Curty, F. Xu, W. Cui, \textit{et al.}, Finite key analysis for measurement-device-independent quantum key distribution. Nat. Commun. \textbf{5}, 3732 (2014).

\bibitem{making} Y.-H. Zhou, Z.-W. Yu, and X.-B. Wang, Making the decoy-state measurement-device-independent quantum key distribution practically useful. Phys. Rev. A \textbf{93}, 042324 (2016).

\bibitem{doublescan} C. Jiang, Z.-W. Yu, X.-L. Hu, \textit{et al.}, Higher key rate of measurement-device-independent quantum key distribution through joint data processing. Phys. Rev. A \textbf{103}, 012402 (2021).

\bibitem{404km} H.-L. Yin, T.-Y. Chen, Z.-W. Yu, \textit{et al.}, Measurement-device-independent quantum key distribution over a 404 km optical fiber. Phys. Rev. Lett. \textbf{117}, 190501 (2016).

\bibitem{442km} J. Y. Liu, X. Ma, H. J. Ding, \textit{et al.}, Experimental demonstration of five-intensity measurement-device-independent quantum key distribution over 442 km. Phys. Rev. A \textbf{108}, 022605 (2023).

\bibitem{HighMDI} K. Wei, W. Li, H. Tan, \textit{et al.}, High-speed measurement-device-independent quantum key distribution with integrated silicon photonics. Phys. Rev. X \textbf{10}, 031030 (2020).

\bibitem{chipMDI} H. Semenenko, P. Sibson, A. Hart, \textit{et al.}, Chip-based measurement-device-independent quantum key distribution. Optica \textbf{7}, 238–242 (2020).

\bibitem{nonstand} G. J. Fan-Yuan, F. Y. Lu, S. Wang, \textit{et al.}, Measurement-device-independent quantum key distribution for nonstandalone networks. Photon. Res. \textbf{9}, 1881-1891 (2021).

\bibitem{PLOB} S. Pirandola, R. Laurenza, C. Ottaviani, \textit{et al.}, Fundamental limits of repeaterless quantum communications. Nat. Commun. \textbf{8}, 15043 (2017).

\bibitem{TF} M. Lucamarini, Z. L. Yuan, J. F. Dynes, \textit{et al.}, Overcoming the rate–distance limit of quantum key distribution without quantum repeaters. Nature (London) \textbf{557}, 400 (2018).

\bibitem{phcodMDI} X. F. Ma, M. Razavi, Alternative schemes for measurement-device-independent quantum key distribution. Phys. Rev. A, 86, 062319 (2012).

\bibitem{SNS} X. B. Wang, Z. W. Yu, and X. L. Hu, Twin-field quantum key distribution with large misalignment error. Phys. Rev. A \textbf{98}, 062323 (2018). 

\bibitem{PM} X. F. Ma, P. Zeng, and H. Y. Zhou, Phase-matching quantum key distribution. Phys. Rev. X \textbf{8}, 031043 (2018). 

\bibitem{NPP} C. H. Cui, Z.-Q. Yin, R. Wang, \textbf{et al.}, Twin-field quantum key distribution without phase postselection. Phys. Rev. Appl. \textbf{11}, 034053 (2019). 

\bibitem{SimpleS} M. Curty, K. Azuma, and H. K. Lo, Simple security proof of twin-field type quantum key distribution protocol. npj Quantum Inf. \textbf{5}, 64 (2019).

\bibitem{WSTF} S. Wang, D.-Y. He, Z.-Q. Yin, \textit{et al.}, Beating the fundamental rate-distance limit in a proof-of-principle quantum Key distribution system. Phys. Rev. X \textbf{9}, 021046 (2019). 

\bibitem{YLiu} Y. Liu, Z.-W. Yu, W. J. Zhang, \textit{et al.}, Experimental twin-field quantum key distribution through sending or not sending. Phys. Rev. Lett. \textbf{123}, 100505 (2019).

\bibitem{HLiu} H. Liu, C. Jiang, H. T. Zhu, \textit{et al.}, Field test of twin-field quantum key distribution through sending-or-not-sending over 428 km. Phys. Rev. Lett. \textbf{126}, 250502 (2021).

\bibitem{duband600} M. Pittaluga, M. Minder, M. Lucamarini, \textit{et al.}, 600-km repeater-like quantum communications with dual-band stabilization. Nat. Photon. \textbf{15}, 530–535 (2021).

\bibitem{830km} S. Wang, Z.-Q. Yin, D.-Y. He, \textit{et al.}, Twin-field quantum key distribution over 830-km fibre. Nat. Photon. \textbf{16}, 154–161 (2022).

\bibitem{1000km} Y. Liu, W.-J. Zhang, C. Jiang, \textit{et al.}, Experimental twin-field quantum key distribution over 1000 km fiber distance. Phys. Rev. Lett. \textbf{130}, 210801 (2023).

\bibitem{local} J. P. Chen, F. Zhou, C Zhang, \textit{et al.}, Twin-Field Quantum Key Distribution with Local Frequency Reference. Phys. Rev. Lett. \textbf{132}, 260802 (2024).

\bibitem{comb546} L. Zhou, J. Lin, C. F. Ge, \textit{et al.}, Independent-optical-frequency-comb-powered 546-km field test of twin-field quantum key distribution. Phys. Rev. Appl. \textbf{22}, 064057 (2024).

\bibitem{AOPP} C. Jiang, Z.-W. Yu, X.-L. Hu and X.-B. Wang, Robust twin-field quantum key distribution through sending or not sending. National Sci. Rev. \textbf{10}, 4 (2023).

\bibitem{YPChen} Y.-P. Chen, J.-Y. Liu, M.-S. Sun, \textit{et al.}, Experimental measurement-device-independent quantum key distribution with the double-scanning method. Opt. Lett. \textbf{46}, 3729–3732 (2021).

\bibitem{MP} P. Zeng, H. Zhou, W. Wu, \textit{et al.}, Mode-pairing quantum key distribution. Nat. Commun. \textbf{13}, 3903 (2022).

\bibitem{MPexp} H.-T. Zhu, Y.-Z. Huang, H. Liu, \textit{et al.}, Experimental Mode-Pairing Measurement-Device-Independent Quantum Key Distribution without Global Phase Locking. Phys. Rev. Lett. \textbf{130}, 030801 (2023).

\bibitem{AMZI} C. Elliott, G. Troxel, Quantum cryptography in practice. In Proceedings of the 2003 conference on Applications, technologies, architectures, and protocols for computer communications (2003), pp. 227-238.

\bibitem{FMI} Z. F. Han, X. F. Mo, Y. Z. Gui, \textit{et al.}, Stability of phase-modulated quantum key distribution systems. Appl. Phys. Lett. \textbf{86}, 22 (2005).
\end{thebibliography}


\section*{Acknowledgments}
We gratefully appreciate Prof. C. F. Li and Prof. Z. L. Yuan for enlightening discussion during the work.
\paragraph*{Funding:}
The Natural  Science Foundation of Jiangsu Province (Grant No. BE2022071), and the National Natural Science Foundation of China (Grant Nos. 12074194, 62471248, 62101285).
\paragraph*{Author contributions:}
J. Y. Liu designed the scheme, performed the experiment, collected and analyzed the data. X. Y. Zhou, H. J. Ding and J. X. Xu helped to perform the experiment, C. H. Zhang assisted in simulations, J. Li assisted in writing, and Q. Wang supervised the whole project.
\paragraph*{Competing interests:}
There are no competing interests to declare.
\paragraph*{Data and materials availability:}
All data needed to evaluate the conclusions of the article are present in the article and / or Supplementary Materials. Additional data related to this article may be requested from the authors.


\subsection*{Supplementary materials}
Supplementary Text\\
Figures S1 to S4\\
Tables S1 to S3\\


\newpage


\renewcommand{\thefigure}{S\arabic{figure}}
\renewcommand{\thetable}{S\arabic{table}}
\renewcommand{\theequation}{S\arabic{equation}}
\renewcommand{\thepage}{S\arabic{page}}
\setcounter{figure}{0}
\setcounter{table}{0}
\setcounter{equation}{0}
\setcounter{page}{1} 


\begin{center}
\section*{Supplementary Materials for\\ \scititle}

Jingyang Liu,
Xingyu Zhou,
Huajian Ding,
Jiaxin Xu,
Chunhui Zhang, \\
Jian Li,
and Qin Wang$^{*}$\\
\small$^\ast$Corresponding author. Email: qinw@njupt.edu.cn\\
\end{center}

\subsubsection*{This PDF file includes:}
Supplementary Text\\
Figures S1 to S4\\
Tables S1 to S3\\

\newpage

\subsection*{Supplementary Text}
\subsubsection*{Measurements of beat note and global phase drifts}
\begin{figure}[ht]
	\centering
	\includegraphics[width=1.0\linewidth]{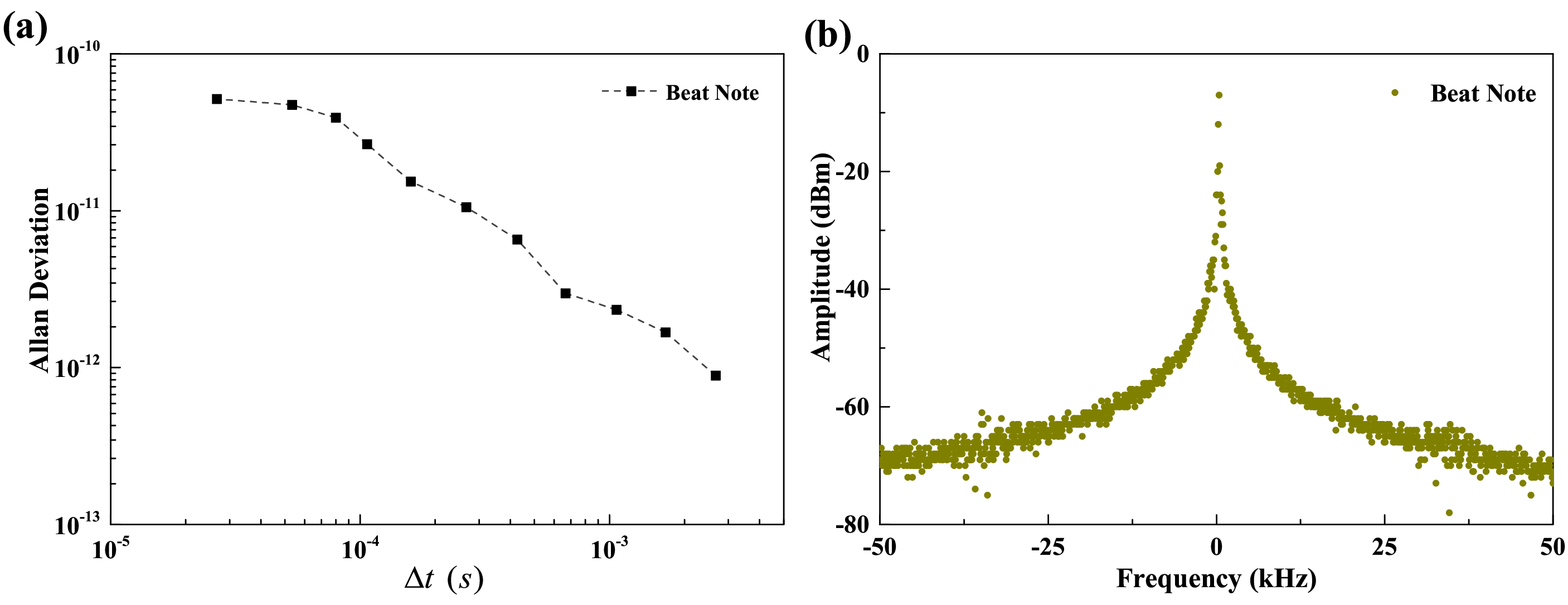}
	\caption{(a) Allan deviation of beat note. (b) Frequency spectrum of locked beams with RBW = 200 Hz.}
	\label{figS1}
\end{figure}

We evaluate the phase-locked beat note in time domain and frequency domain respectively. The measured relative Allan deviation $ 10^{-11}\sim $$ 10^{-13} $ of the beat note in the time domain, shown in Fig. \ref{figS1}a, demonstrates excellent frequency stability. Additionally, the beat note spectra, measured with the resolution bandwidth (RBW) of 200 Hz by a spectrum analyzer, is displayed in Fig. \ref{figS1}b. By using a Lorentz fit, the -3 dB linewidth is about 21 Hz, supposing an aligned optical frequency between Alice and Bob's laser. In current setting, the residual mean-square phase error associated with the OPLL is approximately $ 5.6\times10^{-3}\ \mathrm{rad^2} $. 

\begin{figure}[!htb]
	\centering
	\includegraphics[width = 1.0\linewidth]{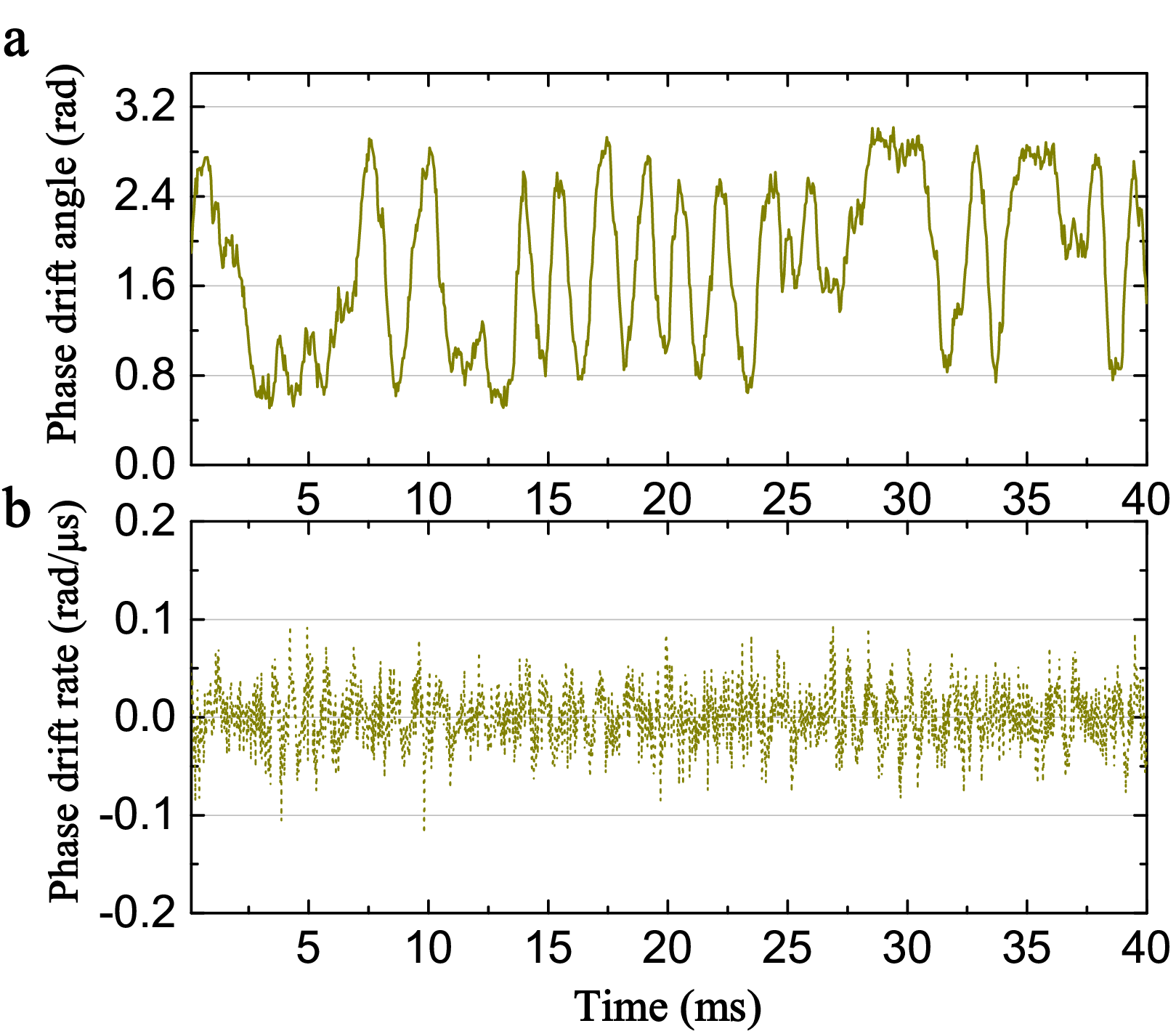}
	\caption{Phase drift at 300 km. The maximum drift rate is 6.34 $ rad ms^{-1} $.}
	\label{FigS3}
\end{figure}

\begin{figure}[!htb]
	\centering
	\includegraphics[width = 1.0\linewidth]{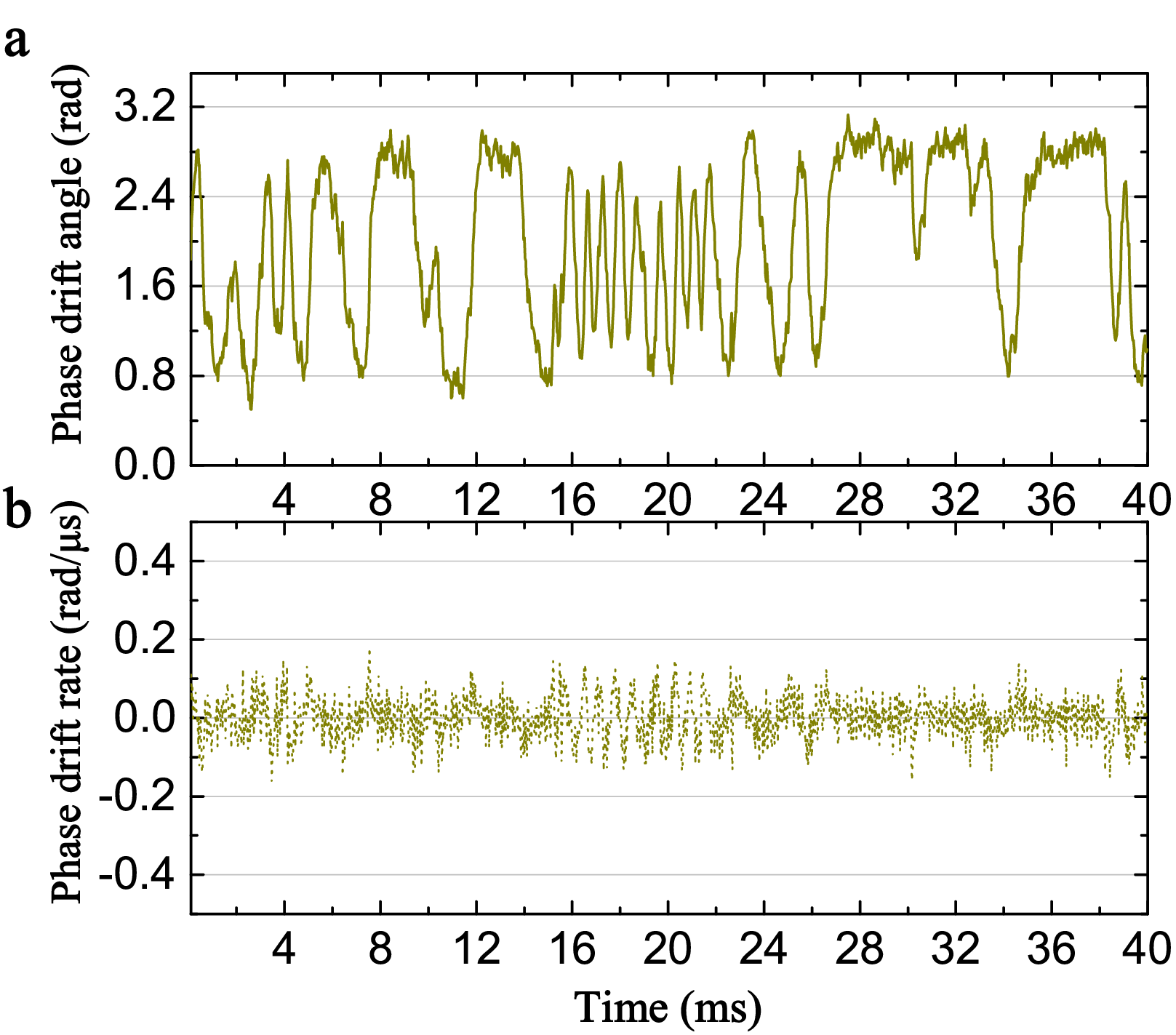}
	\caption{Phase drift at 400 km. The maximum drift rate is 14.95 $ rad ms^{-1} $.}
	\label{FigS4}
\end{figure}

And the measured global phase drift angle and the phase drift rate from 0 km to 500 km are shown in Fig. \ref{FigS3} to \ref{FigS4} respectively. Ensuring the validity, Charlie's beam splitter output is connected to the photoelectric probe (5 GHz bandwidth), which links to an oscilloscope.

\subsubsection*{Two-phase scan}
\begin{figure}[!htb]
	\centering
	\includegraphics[width = 0.8\linewidth]{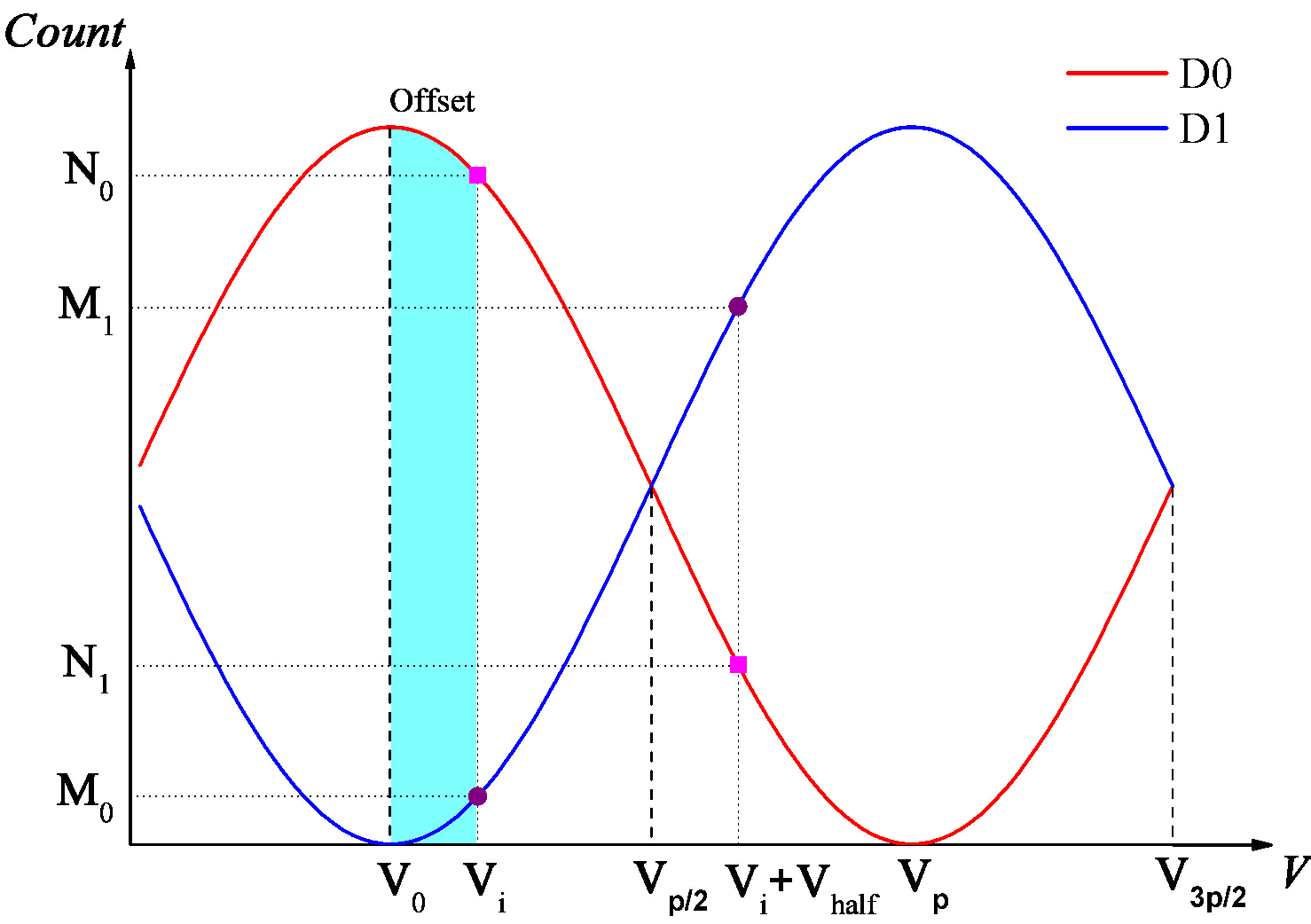}
	\caption{Schematic of two-phase scan (2PS).}
	\label{FigS2}
\end{figure}

In this section, we give out the details of two-phase scan (2PS) estimation method. Here brighter reference lights sent from Alice and Bob are used to observe single-photon interference at Charlie side. The output intensity of the BS can be written as:
\begin{align}\label{I}
	I=\dfrac{I_{0}}{2}[1+cos(\phi)],
\end{align}
where $ \phi $ is the total phase difference between Alice and Bob, defined as $ \phi=\vartheta_{A}-\vartheta_{B}+\Delta\varphi $. $ \vartheta_{A} $ ($ \vartheta_{B} $) denotes Alice's (Bob's) phase modulation. $ \Delta\varphi $ is the phase fluctuation caused by wavelength difference and phase drift of channel. Here, we assume that the wavelength difference is suppressed to a very low constant which can be realized via a optical phase-locked loop. Connecting the first output to the detector D0, we can rewrite Eq.(\ref{I}) as:
\begin{align}\label{cD0}
	C=C_{0}cos(\phi)+C_{0}+C_{d},
\end{align}
where $ C_{0} $ denotes the amplitude of interference and $ C_{d} $ represents the count floor, contributed from the interference of encoded pulses and detector's dark counts. Hence, the second output of BS is:
\begin{align}\label{cD1}
	C=C_{0}cos(\phi+\pi)+C_{0}+C_{d}.
\end{align}
With the chopper and detector window, $ C_{d}\approx0 $ in a short period. Thus, we can plot the interference output of two detectors with respect to the PM's offset voltage in Fig. \ref{FigS2}. $ V_{0} $ is the zero-phase voltage which can be efficiently estimated via 2PS.

During two-phase scanning, Alice's and Bob's PM voltage are all fixed to 0 V, we first apply an arbitrary initial voltage $ V_{i} $ to the PM at Charlie for 20 $ \mu s $, and then apply $ V_{i}+V_{half}/2 $ to it for another 20 $ \mu s $. $ V_{half} $ denotes the PM's half wave voltage. Accordingly, the recording counts during the first period of D0 (D1) is $ N_{0} $ ($ M_{0} $), and counts during the second one of D0 (D1) is $ N_{1} $ ($ M_{1} $). These four counts form the count matrix. Based on Eq.(\ref{cD0}) and (\ref{cD1}), we get,
\begin{align}
	&\dfrac{N_{0}}{N_{0}+M_{0}}=\dfrac{1+cos(\dfrac{V_{c}\pi}{V_{half}})}{2}, \label{cos} \\
	&\dfrac{N_{1}}{N_{1}+M_{1}}=\dfrac{1-sin(\dfrac{V_{s}\pi}{V_{half}})}{2}, \label{sin}
\end{align}
where $ V_{c} $ ($ V_{s} $) denotes the offset from zero-phase voltage calculated by $ arccos $ ($ arcsin $). Thus, the estimated zero-phase voltage calculated from $ arccos $ is 
\begin{equation}
	V'=
	\begin{cases}
		V_{i}-V_{c}, \:\:\: if \: 1-2N_{1}/(N_{1}+M_{1})>0; \\
		V_{i}+V_{c}, \:\:\: if \: 1-2N_{1}/(N_{1}+M_{1})\leq0 .
	\end{cases}
\end{equation}
And the zero-phase voltage calculated from $ arcsin $ is 
\begin{equation}
	V''=
	\begin{cases}
		V_{i}-V_{s}, \:\:\: if \: 2N_{0}/(N_{0}+M_{0})-1>0; \\
		V_{i}+V_{s}-V_{half}, \:\:\: if \: 2N_{0}/(N_{0}+M_{0})-1\leq0 .
	\end{cases}
\end{equation}
Generally, $ V'=V'' $ if there exists no shot-noise and systematic errors. However, these two problems are non-negligible under practical circumstances, which makes a deviation between $ V' $ and $ V'' $. We then take the average of them as the final estimated zero-phase voltage, and obtain the corresponding phase slice number calculated proportionally.

\clearpage

\clearpage

\begin{table}
	\centering
	\caption{Optimized parameters of MDI-QKD and SNS-QKD protocols.}
	\begin{tabular}{p{1.5cm}p{1.5cm}p{1.5cm}p{1.5cm}p{1.5cm}p{1.5cm}p{1.5cm}p{1.5cm}}
		\hline
		\hline
		Distance  & $ 150 $ km & $ 241  $ km & $ 241   $ km & $ 310 $ km  & $ 351 $ km  & $ 400 $ km  & $ 431 $ km  \\
		Protocol & MDI & MDI & TF & TF & TF & TF & TF \\
		\hline
		$ \mu $ & $ 0.7681 $ & $ 0.6853 $  & $ 0.5012 $ & $ 0.4995  $ & $ 0.4973 $ & $ 0.4908 $ & $ 0.4890 $ \\
		$ \nu $ & $ 0.2601 $ & $ 0.3136 $  & $ 0.4917 $ & $ 0.4920  $ & $ 0.4972 $ & $ 0.4907 $ & $ 0.4889 $ \\
		$ \omega $ & $ 0.0730 $  & $ 0.0742  $ & $ 0.0187  $ & $ 0.0238 $ & $ 0.0459 $ & $ 0.0518 $ & $ 0.0359 $ \\
		$ o $ & $ 0.001 $  & $ 0.001  $ & $ 0.001 $ & $ 0.001 $ & $ 0.001 $ & $ 0.001 $ & $ 0.001 $ \\
		$ p_\mu $ & $ 0.6041 $ & $ 0.4453 $  & $ 0.9452 $ & $ 0.9221  $ & $ 0.8136 $ & $ 0.7575 $ & $ 0.8583 $ \\
		$ p_\nu $ & $ 0.1028 $ & $ 0.1372 $  & $ 0.0011 $ & $ 0.0026  $ & $ 0.0077 $ & $ 0.0103 $ & $ 0.0052 $ \\
		$ p_\omega $ & $ 0.2745 $  & $ 0.3994 $ & $ 0.0331  $ & $ 0.0467 $ & $ 0.1241 $ & $ 0.1583 $ & $ 0.0776 $ \\
		$ p_o $ & $ 0.01866 $  & $ 0.01818  $ & $ 0.0206  $ & $ 0.0286 $ & $ 0.0545 $ & $ 0.0739 $ & $ 0.0588 $ \\
		$ \epsilon $ & /  & / & $ 0.2809 $  & $ 0.2808 $ & $ 0.2808 $ & $  0.2807 $ & $ 0.2808 $ \\
		\hline
		\hline
	\end{tabular}
	\label{tabS1}
\end{table}

\begin{table}[htb]
	\centering
	\caption{Experimental results at various fiber distances with the SNS-QKD protocol.}
	\begin{tabular}{p{3.5cm}p{2.2cm}p{2.2cm}p{2.2cm}p{2.2cm}p{2.2cm}}
		\hline
		\hline
		Distance & $ 241   $ km & $ 310 $ km  & $ 351 $ km  & $ 400 $ km  & $ 431 $ km \\
		Loss & $ 40.95 $ dB & $ 52.77 $ dB & $ 59.68 $ dB & $ 68.02 $ dB & $ 73.29 $ dB \\
		\textit{N} & $ 1.0\times10^{12} $ & $ 1.0\times10^{12}  $ & $ 1.0\times10^{11}   $ & $ 1.0\times10^{11} $ & $ 1.0\times10^{12} $ \\
		\hline
		Detected $ \mu\mu $ & $ 402944070 $  & $ 99251388  $ & $ 3449894   $ & $ 1131423 $ & $ 7894468 $  \\
		Detected $ \mu0 $ & $ 516544784 $  & $ 127151740  $ & $ 4419713   $ & $ 1449871 $ & $ 10114509 $  \\
		Detected $ 0\mu $ & $ 516544821 $  & $ 127151460  $ & $ 4419480   $ & $ 1450431 $ & $ 10114081 $  \\
		Detected $ 00 $ & $ 2776 $  & $ 2634  $ & $ 205   $ & $ 178 $ & $ 1188 $  \\
		Detected $ \nu\nu $ & $ 1277 $  & $ 951  $ & $ 552   $ & $ 360 $ & $ 502 $  \\
		Detected $ \nu o $ & $ 66430 $  & $ 55134  $ & $ 13953   $ & $ 9575 $ & $ 21024 $  \\
		Detected $ o \nu $ & $ 66271 $  & $ 54922  $ & $ 14076   $ & $ 9591 $ & $ 20729 $ \\
		Detected $ \omega\omega $ & $ 29301 $  & $ 19313  $ & $ 13092   $ & $ 8696 $ & $ 7652 $ \\
		Detected $ \omega o $ & $ 72885 $  & $ 47246  $ & $ 20752   $ & $ 15550 $ & $ 23166 $  \\
		Detected $ o \omega $ & $ 72966 $  & $ 47850  $ & $ 20591   $ & $ 15436 $ & $ 22855 $ \\
		Detected $ oo $ & $ 105 $  & $ 126  $ & $ 18   $ & $ 22 $ & $ 57 $ \\
		QBER of $ \nu\nu $ & $ 4.54\% $  & $ 4.71\%  $ & $ 4.80\%   $ & $ 4.91\% $ & $ 5.10\% $  \\
		QBER of $ \omega\omega $ & $ 4.69\% $  & $ 4.87\%  $ & $ 4.98\%   $ & $ 5.08\% $ & $ 5.22\% $  \\
		$ n_{t} $ (after AOPP) & $  290193568 $  & $ 71422087  $ & $ 2483070  $ & $ 815165 $ & $ 5690680 $  \\
		$ E_{Z} $ (before AOPP) & $ 28.01\% $  & $ 28.06\%  $ & $ 28.10\%   $ & $ 28.17\% $ & $ 28.19\% $  \\
		$ E_{Z} $ (after AOPP) & $ 0.048\% $  & $ 0.16\%  $ & $ 0.36\%  $ & $ 0.95\% $ & $ 1.06\% $  \\
		$ n_{11}^{Z} $ (before AOPP) & $ 574237169 $  & $ 139619399  $ & $ 4637054  $ & $ 1495854 $ & $ 10819103 $ \\
		$ n_{11}^{Z} $ (after AOPP) & $ 89418924 $  & $ 21409770  $ & $ 662790  $ & $ 205423 $ & $ 1593451 $  \\
		$ e_{ph} $ (before AOPP) & $ 6.21\% $  & $ 6.83\%  $ & $ 8.57\%  $ & $ 9.28\% $ & $ 8.54\% $  \\
		$ e_{ph} $ (after AOPP) & $ 11.72\% $  & $ 11.87\%  $ & $ 16.64\%  $ & $ 18.66\% $ & $ 16.23\% $  \\
		$ R\:(bit/pulse) $ & $ 4.26\times10^{-5} $  & $ 9.37\times10^{-6}  $ & $ 2.20\times10^{-6}   $ & $ 5.29\times10^{-7} $ & $ 4.57\times10^{-7} $  \\
		$ R\:(bps) $ & $ 2105.59 $  & $ 463.52  $ & $ 108.91   $ & $ 26.16 $ & $ 22.60 $  \\
		\hline
		\hline
	\end{tabular}
	\label{tabS2}
\end{table}

\begin{table}[htb]
	\centering
	\caption{Experimental results at various fiber distances with the MDI-QKD protocol.}
	\begin{tabular}{p{4.0cm}p{4.0cm}p{4.0cm}}
		\hline
		\hline
		Distance & $ 150   $ km & $ 241    $ km  \\
		Loss     & $ 25.50 $ dB & $ 40.95  $ dB  \\
		\textit{N} & $ 1.0\times10^{11} $ & $ 10^{12} $ \\
		\hline
		Detected $ \mu\mu $ & $ 5888709 $ & $ 750413  $  \\
		Detected $ \mu o $ & $ 90344 $ & $ 15074  $  \\
		Detected $ o \mu $ & $ 90216 $ & $ 15132  $  \\
		Detected $ \nu\nu $ & $ 40738 $ & $ 31004  $  \\
		Detected $ \nu o $ & $ 1851 $ & $ 990  $  \\
		Detected $ o \nu $ & $ 1973 $ & $ 1003  $  \\
		Detected $ \omega\omega $ & $ 23184 $ & $ 14395  $  \\
		Detected $ \omega o $ & $ 405 $ & $ 174  $  \\
		Detected $ o \omega $ & $ 419 $ & $ 207  $  \\
		Detected $ oo $ & $ 1 $ & $ 3  $  \\
		QBER of $ \mu\mu $ & $ 0.12\% $ & $ 0.15\%  $  \\
		QBER of $ \nu\nu $ & $ 25.44\% $ & $ 25.55\%  $  \\
		QBER of $ \omega\omega $ & $ 25.49\% $ & $ 25.70\%  $  \\
		$ n_{11}^{Z} $ & $ 3637719 $ & $ 834258 $  \\
		$ e_{ph} $ & $ 22.55\% $ & $ 26.10\% $  \\
		$ R\:(bit/pulse) $ & $ 2.17\times10^{-6} $ & $ 1.46\times10^{-8}  $  \\
		$ R\:(bps) $ & $ 107.61 $ & $ 0.72  $  \\
		\hline
		\hline
	\end{tabular}
	\label{tabS3}
\end{table}

\end{document}